\documentclass[11pt]{article}

\usepackage[utf8]{inputenc}
\usepackage[margin=2.5cm]{geometry}
\usepackage{graphicx}
\usepackage{amsmath,amssymb}
\usepackage{booktabs}
\usepackage{siunitx}
\usepackage{multirow}
\usepackage{caption}
\usepackage{subcaption}
\usepackage{hyperref}
\usepackage{authblk}
\usepackage{lineno}
\usepackage{setspace}

\hypersetup{
    colorlinks=true,
    citecolor=blue,
    linkcolor=blue,
    urlcolor=blue
}

\title{Electronic structure of a persistent excitonic fluid far beyond the Mott limit}

\author[1,2]{Weiyan Qi}
\author[1]{Stefano Ponzoni}
\author[1,3]{Ashim Pramanik}
\author[1]{Vitaly Gorelov}
\author[1]{Yannis Laplace}
\author[4,5]{Gianluca Stefanucci}
\author[4,5]{Enrico Perfetto}
\author[4,5]{Esteban Morcos}
\author[2]{Marino Marsi}
\author[2]{Steven Francis Udaundo}
\author[6,7,8]{Sebastian Maherlein}
\author[9]{Amina Taleb-Ibrahimi}
\author[10,11]{Kaibo Zheng}
\author[10]{T\"onu Pullerits}
\author[1,*]{Luca Perfetti}

\affil[1]{Laboratoire des Solides Irradi\'es, CEA/DRF/lRAMIS, CNRS, Ecole Polytechnique, Institut Polytechnique de Paris, Palaiseau, France}
\affil[2]{Laboratoire de Physique des Solides, Université Paris-Saclay, CNRS, 91405 Orsay, France}
\affil[3]{Department of Physics, School of Sciences and Humanities, SR University, Warangal, Telangana, 506 371, India.}
\affil[4]{Physics Department, University of Rome Tor Vergata, Rome, Italy}
\affil[5]{INFN, Sezione di Roma Tor Vergata, Rome, Italy}
\affil[6]{Department of Physical Chemistry, Fritz Haber Institute of the Max Planck Society, 14195, Berlin, Germany}
\affil[7]{Institute of Radiation Physics, Helmholtz-Zentrum Dresden-Rossendorf, 01328, Dresden, Germany}
\affil[8]{Institute of Applied Physics, Dresden University of Technology, 01187, Dresden, Germany}
\affil[9]{Soci\'{e}t\'{e} civile Synchrotron SOLEIL, L'Orme des Merisiers, Saint-Aubin - BP 48, 91192 GIF-sur-YVETTE, France}
\affil[10]{Department of Chemistry, Technical University of Denmark, DK-2800 Kongens Lyngby, Denmark}
\affil[11]{Chemical Physics and NanoLund, Lund University, Box 124, 22100 Lund, Sweden}

\affil[*]{Correspondence: luca.perfetti@polytechnique.edu}
\date{}

\begin{document}

\maketitle

\begin{abstract}
Excitons are conventionally expected to destabilize at high photoexcitation densities, as screening weakens the electron–hole attraction. Recent studies, however, challenge this paradigm, suggesting that excitons generated by strong optical pumping can remain stable even beyond the onset of population inversion. Here, using time- and angle-resolved photoelectron spectroscopy on 2D hybrid perovskites, we provide direct evidence for such a high-density excitonic state, characterized by a distinctive Mexican-hat dispersion. Remarkably, this excitonic band reconstruction endures for orders of magnitude longer than the optical coherence time, revealing a population-sustained electronic structure that survives well into the incoherent regime. The excitonic resonance also persists far beyond optical saturation, at densities where exciton wavefunctions strongly overlap. Our findings reveal a long-lived, high-density quantum fluid of excitons with characteristics reminiscent of a BCS condensate, an electronic structure markedly different from that of the dilute limit.

\end{abstract}

\section*{Main}

In semiconductors, Coulomb interactions bind electrons and holes into excitons, giving rise to sharp absorption peaks below the electronic bandgap and efficient radiative recombination. In the dilute regime, excitons behave as well-defined, charge-neutral quasiparticles. They have been extensively investigated, particularly in two-dimensional materials, where reduced dielectric screening strongly enhances their binding energy and oscillator strength \cite{Blancon,Deng}.

The fate of excitons at high densities remains debated. In the conventional Mott picture, increasing carrier density enhances screening and eventually destabilizes excitons, leading to their ionization into unbound electron–hole pairs \cite{Asano,Shimano}. The optical response changes dramatically as the system reaches population inversion: the exciton absorption peak progressively disappears while optical gain emerges \cite{Heinz}. THz spectroscopy has also revealed the coexistence of bound excitons and unbound carriers across the Mott transition \cite{Heinz}. This exciton–plasma crossover is gradual and depends on exciton properties, including dimensionality, binding energy and Bohr radius. In semiconductors with strongly anisotropic screening, excitonic signatures can persist up to photoexcited carrier densities of ($10^{14}~\mathrm{cm}^{-2}$) \cite{Heinz,Baldini}.

However,the conventional Mott picture assumes thermally distributed carrier populations, which need not always be realized following photoexcitation. For nonthermal distributions, such as those generated by resonant pumping at the optical gap, theory predicts that excitons can persist to remarkably high carrier densities through the emergence of exceptional points in the spectrum of the non-Hermitian Bethe–Salpeter Hamiltonian \cite{Stefanucci,Perfetto}. This topological mechanism gives rise to a quantum fluid with spectral features reminiscent of a BCS condensate, in which bound electron–hole pairs remain protected against ionization even when embedded in the electron-hole continuum.

In Time- and angle-resolved photoelectron spectroscopy (tr-ARPES), the quantum exciton fluid at high density is predicted to manifest itself through a characteristic band reconstruction with a Mexican-hat dispersion \cite{Stefanucci,Perfetto}. Such a dispersion has so far been observed only in the coherent regime and interpreted in terms of many-body Floquet dressing \cite{Dani3}. Within this picture, the reconstructed bands are expected to disappear as the optical coherence decays. However, theory based on T-matrix diagrams predicts that the excitonic band reconstruction can survive well beyond this timescale, sustained by the exciton population rather than the macroscopic polarization \cite{Perfetto}.

Here, we investigate this long-living state using tr-ARPES \cite{Qi,Cherasse} on the two-dimensional hybrid perovskite  (BA)$_2$(MA)$_2$Pb$_3$I$_{10}$, a Ruddlesden–Popper compound in which inorganic quantum wells are separated by organic layers \cite{Stoumpos,Liang} (Fig. 1a). The reduced dimensionality stabilizes bound electron–hole states up to room temperature, making this material a suitable platform for investigating high-density excitonic phenomena \cite{Stoumpos,Deng,Blancon}.
Beyond resolving their dispersion, tr-ARPES enables the tomographic reconstruction of excitonic wavefunctions and has emerged as a powerful probe of excitons in two-dimensional (2D) transition metal dichalcogenides \cite{Dani1}, organic crystals \cite{Ernstorfer,Hoefer1}, topological insulators \cite{Lanzara}, moiré heterostructures \cite{Mathias,Dani2} and hybrid structures \cite{Bennecke}.

By mapping the photoelectron distribution over a wide range of resonant pump fluences, we reveal a characteristic Mexican-hat band reconstruction that survives at excitation fluences three times the optical saturation threshold and persists well beyond the optical coherence time. These findings provide evidence for a stable quantum exciton fluid in which bound electron–hole correlations remain robust against screening-induced ionization.

Figure 1 shows tr-ARPES data acquired on a (BA)$_2$(MA)$_2$Pb$_3$I$_{10}$ crystal at 50 K. Energy- and momentum-resolved ARPES maps \cite{Faure} are first recorded at very early delays (50 fs for panel c-d and 100 fs for panel e) using a weak pump fluence ($\sim 3~\mu$J/cm$^2$), for three representative pump photon energies. As shown schematically in Fig. 1b, optical transitions at 2.4 eV, 2.22 eV and 2.07 eV lie above the conduction band minimum just below it, and at the exciton resonance, respectively. Excitation with 2.4 eV photons generates free carriers in the conduction band (dashed white line in Fig. 1c), along with a weak spectral signature at the exciton resonance. At 2.22 eV (Fig. 1d), the response is dominated by population at the conduction band minimum (CBM), with only a faint excitonic contribution. In contrast, resonant pumping at 2.07 eV selectively prepares a coherent exciton population (Fig. 1e). The energy separation between the CBM and the exciton resonance yields an exciton binding energy of $E_B=0.18$ eV, in excellent agreement with prior all-optical measurements \cite{Blancon}. Notice in Fig. 1e, that the spectral weight is strongly peaked at the center of the Brillouin zone and exhibits the characteristic downward dispersion previously reported in transition-metal dichalcogenides \cite{Dani2,Man2021}. As theoretically expected at low excitation density \cite{Perfetto}, the overall shape of the ARPES intensity map closely follows the curvature of the calculated valence-band replica.

The wavevector distribution measured by tr-ARPES in Fig. 1e (weak and resonant excitation) directly reflects the squared modulus of the exciton wavefunction envelope in reciprocal space \cite{Perfetto2016}. Experimentally, this quantity (Fig. 1f) is extracted from a Momentum Distribution Curve (MDC) by integrating the ARPES intensity in Fig. 1e over energy. Theoretically, the wavefunction is extracted from Bethe–Salpeter calculations based on a tight-binding parametrization of the two-dimensional perovskite multiple-quantum-well structure \cite{Stoumpos}. The tight-binding model is constructed by fitting the conduction and valence bands to the \emph{ab initio} electronic band structure. This approach yields an imaginary part of the dielectric function, which well agrees with the exciton resonance observed in the experimental optical absorption spectrum and predicts an exciton binding energy of $E_B=0.186$ eV. Moreover, as shown in Fig. 1f, the calculated envelope of the exciton wavefunction in wavevector space ($|\psi (k)|^2$) overlaps with the experimental MDC. From this analysis, we infer in-plane root-mean-square radius of $\sqrt{\langle r^2 \rangle}=4$ nm, confirming the generation of a spatially delocalized Wannier excitons (Fig. 1g).

Being bound electron–hole pairs, bright excitons possess a large transition dipole moment. To demonstrate this directly, we monitor the integrated tr-ARPES intensity while tuning the pump photon energy across the absorption edge. The resulting excitation density exhibits a pronounced resonance at 2.07 eV (Fig. 2a), consistent with the optical absorption spectrum and photoluminescence measured over the same spectral range (Fig. 2b). Figure 2c shows the integrated photoelectron intensity as a function of pump fluence, measured under resonant excitation and 100 fs after the arrival of the pump pulse. The approximately linear dependence observed up to 50 $\mu$J/cm$^2$ allows us to calibrate the tr-ARPES intensity against the photoexcitation density, assuming an optical penetration depth of 140 nm. Above this fluence, however, the photoexcitation density deviates from the linear scaling and gradually approaches a saturation value of $5 \times 10^{12}$ cm$^{-2}$. Remarkably, the pronounced excitonic resonance persists well beyond the fluence $F_s=100$ $\mu$J/cm$^2$ at which the photoexcitation density reachs 75\% of the saturation threshold. As the pump fluence increases to $F_s$ (Fig. 2d), the resonance broadens only modestly. Even at 350 $\mu$J/cm$^2$, corresponding to $3.5 F_s$, the resonance remains clearly visible (Fig. 2e). Thus, a bound state with a large transition dipole moment can still be efficiently excited under these extreme driving conditions. Such a behavior contrasts with the conventional Mott picture, which predicts that increasing carrier density progressively screens the Coulomb interaction and ultimately drives the dissociation of excitons into unbound electron–hole pairs.

The crossover from weak to high photoexcitation density is monitored by acquiring ARPES intensity maps as a function of an increasing pump fluence. We present data for exciton densities in the range $1.9\times 10^{11}$ to $5\times 10^{12}$/cm$^2$ and acquired at pump probe delay of 100 fs. At the highest photoexcitation density, approximately 2.5 electron–hole pairs are confined within the area defined by the two-dimensional exciton wavefunction, $\pi\langle r^2\rangle$, placing the system deep in the quantum regime. The ARPES map of the exciton gas (low density) differs markedly from that of the quantum exciton fluid (high density). Whereas the spectral weight disperses downward in the dilute regime (Fig. 3a), an upward dispersion emerges in the quantum fluid and becomes clearly visible at an excitation density of $1.9 \times 10^{12}$ cm$^{-2}$ (Fig. 3b). At even higher densities, the upward dispersion becomes more pronounced, approaching that of electrons in the conduction band (white dashed line in Fig. 3c-d).

In the high fluence regime, the remnant hybridization with the valence band replica shapes the faint spectral weight at large momenta. To aid visualization, the tr-ARPES spectra in Fig. 3e–h are obtained normalizing Energy Distribution Curves (EDCs) to their maximum intensity. At intermediate fluences (Fig. 3f,g), the resulting maps reveal a characteristic Mexican-hat-like dispersion, reminiscent of that expected for a steady-state exciton liquid in a BCS-like superfluid regime. Upon further increasing the carrier density (Fig. 3h), the shadow replica of the valence band shifts beyond the momentum window of $[-0.1, 0.1]$ Å$^{-1}$. Out of this momentum range, the signal becomes too weak to be resolved above the noise level, even after normalization.

We simulate the experimental results using the TR-ARPES formula developed in Ref. \cite{Perfetto}. The nonequilibrium excitonic spectrum entering the formula is calculated within the exceptional-exciton framework \cite{Stefanucci}, going beyond the first-order expansion in excitation density adopted in Ref \cite{Perfetto}. Specifically, exciton energies and wavefunctions are obtained by diagonalizing the pseudo-Hermitian Bethe–Salpeter Hamiltonian (PH-BSH) for electronic populations characteristic of resonant pumping, rather than the thermal Fermi–Dirac distribution assumed in the conventional Mott picture. This choice is justified by the slow transfer from the resonantly pumped bright exciton to the lowest-energy spin-dark exciton, which occurs on a timescale of approximately 1 ns \cite{Manfred}.

Consequently, the carrier distribution retains the characteristic profile of resonant pumping even a few picoseconds after excitation, when the system has entered the incoherent regime. A key consequence of these distributions is that the entire PH-BSH spectrum remains real up to excitation densities at which population inversion occurs (in our material near the $\Gamma$ point).
At the critical inversion density, the conventional exciton evolves into an exceptional point corresponding to a bound state embedded in the electron–hole continuum.  This topological property provides a natural protection mechanism against hybridization with the continuum, offering a physical basis for the persistence of bound excitons at these high excitation densities. As shown in Fig. 3(i-p), the simulated tr-ARPES spectra are in close agreement with the experimental data. At low fluence, spectral weight can be understood as a sideband replica of the valence band, with a momentum dependent weight given by the squared excitonic wavefunction, $|\psi(k)|^2$. Instead, when the photoexcitation density is higher than the critical value, the hybridization between conduction band and valence-band replicas gives rise to the characteristic Mexican-hat–like structure.


As shown in Fig. 4a, the dynamics of the quantum electron fluid reveal two clearly distinguishable timescales. The integrated tr-ARPES intensity initially decreases by approximately 10\% on a timescale of ~100 fs, followed by a much slower decay with a characteristic timescale of ~20 ps. The initial femtosecond dynamics occurs on a timescale comparable to the loss of optical coherence measured in 2D perovskites by four-wave mixing \cite{Silva}. Notably, the ultrafast component emerges only at pump fluences between 25 and 30 $\mu$J/cm$^2$ (Fig. 4b), coinciding with the onset of the Mexican-hat dispersion (Fig. 3b). Thus, if the fast dynamics originates from decoherence, its signature in the ARPES signal appears to become measurable only above the critical density required to form a BCS-like fluid.

Importantly, the ultrafast dynamics primarily affects the overall photoelectron intensity, while the spectral-weight distribution and its energy–momentum dispersion remain largely unchanged (Fig. 4c-d). These features are instead governed by the exciton population, which decays predominantly through recombination processes on a fluence-dependent timescale approximately two orders of magnitude longer than that of the decoherence. Figure 4c–h shows the corresponding ARPES intensity maps following an initial photoexcitation density of $3.2 \times 10^{12}$ cm$^{-2}$. The upward dispersion characteristic of the quantum exciton liquid persists for up to ~10 ps. By ~30 ps, the exciton density has fallen below the critical point, and the excitonic states are no longer embedded in the particle–hole continuum.

In conclusion, we provide experimental evidence for the formation of a high-density quantum exciton fluid in a two-dimensional hybrid perovskite. At the high excitation densities required to reach this state, the characteristic Mexican-hat dispersion represents a substantial reconstruction of the electronic structure. Remarkably, this reconstruction persists for several picoseconds, deep into the incoherent regime, demonstrating that exciton populations can reshape the electronic structure of a material well beyond the lifetime of the macroscopic polarization \cite{Dani3,Gedik,Zhou,Monney}. Sustaining a comparably strong light-induced reconstruction over such timescales would require intense and prolonged optical excitation, with the inevitable consequence of sample damage. 

The observation of stable excitons at densities exceeding the conventional Mott limit, extending into the population-inverted regime and up to the saturation density, challenges the established picture of exciton ionization at high carrier densities and supports recent theoretical predictions based on topological protection against hybridization with the electron–hole continuum. These findings highlight internal excitations as a route for controlling electronic properties of quantum materials.

\section*{Methods}

Time resolved ARPES experiments have been acquired with a probe beam energy of 4.7 eV and a variable pump energy provided by optical parametric amplifier. In all reported data the sample temperature is kept at 50 K. Both probe and pump photon beams are $P$ polarized. The cross correlation between pump and probe beam has Full width half maximum of $\sim 100$ fs. Single crystals of (BA)$_2$(MA)$_2$Pb$_3$I$_{10}$ have been cleaved at room temperature and base pressure of $2\times 10^{-10}$ mbar. Tr-ARPES maps have been collected along the $\Gamma-X$ direction by an electrostatic spectrometer. We estimate an overall energy resolution of 60 meV and wavevector resolution of 0.1 nm$^{-1}$.

The electronic structure of a single inorganic Pb--I layer is described by the symmetry-based $sp^3$ tight-binding model
of Boyer-Richard et al. \cite{ChoBerkelbach2019}, with one Pb $(s,p)$ and three I $(s,p)$ sites per unit cell. Exciton energies, binding energies, radii, and absorption line shapes have been computed from the Bethe--Salpeter equation (BSE) scheme, following the approach of Cho and Berkelbach \cite{ChoBerkelbach2019} for layered hybrid perovskites. The agreement with available data supports the accuracy of the modelling. In particular, we find that the lowest-energy excited state is a twofold-degenerate spin-dark exciton, lying 20 meV below the first bright exciton, which is also twofold degenerate. Each of these four excitons involves a single conduction band and a single valence band, with distinct band combinations for the four excitons. Since the distinction is primarily due to spin, their spatial wavefunctions are essentially identical. The single-band character of the populated exciton allows us to simulate the nonequilibrium ARPES spectrum using the
formula \cite{Stefanucci,Perfetto} $I(k,\omega) \propto f_k \delta(\omega-E_k)$, where $f_k$ is the population of electrons in the
conduction band after resonant excitation and $E_k$ is the hybridized replica of the valence band. Both quantities can
be expressed in terms of the nonequilibrium valence and conduction dispersions, $\epsilon_{v/c}$, and the corresponding
valence--conduction excitonic coupling, $\delta\Sigma_X \propto \sqrt{n}$, where $n$ is the excitation
density. As shown in Ref.~\cite{Stefanucci},
\begin{equation}
    f_k =
    \frac{E_X+\epsilon_{v k}-E_k}
    {\sqrt{
        (\epsilon_{c k}-E_X-\epsilon_{v k})^2
        +4|\delta\Sigma_{X k}|^2
    }}
    \label{eq:fk}
\end{equation}
and
\begin{equation}
    E_k = \frac{1}{2}
    \left[
        \epsilon_{c k}+E_X+\epsilon_{v k}
        -\sqrt{
            (\epsilon_{c k}-E_X-\epsilon_{v k})^2
            +4|\delta\Sigma_{X k}|^2
        }
    \right].
    \label{eq:Ek}
\end{equation}

In these expressions, the nonequilibrium exciton energy $E_X$ is the lowest-energy discrete eigenvalue of the BSE Hamiltonian for $n<n_c$. At the critical density $n_c$, this eigenvalue merges into the electron--hole continuum and becomes an exceptional point for $n>n_c$. Remarkably, the corresponding eigenfunction remains localized in real space. These nontrivial topological solutions, known as exceptional excitons, underlie the characteristic Mexican-hat band
reconstruction.

\section*{Data availability}

The data supporting the findings of this study are available from the
corresponding author upon reasonable request.

\section*{Code availability}

Custom analysis code is available from the corresponding author upon
reasonable request.

\section*{Acknowledgements}

We acknowledge financial support of the 2D-HYPE project from the Agence Nationale de la Recherche (ANR, Nr. ANR 26-CE30-7340), of the R\'egion Ile de France via (FemtoARPES2.0 project N. EX079194), of the INFN through project TIME2QUEST and of the Horizon Europe research and innovation program of the European Union under the Marie Sklodowska-Curie Grant Agreement No. 101118915 (TIMES). Weiyan Qi acknowledges the China Scholarship Council (CSC) for funding his PhD. Computational resources were provided by GENCI (Project No. 544).

\section*{Author contributions}

W.Q., S.P., A.P. and S.F.U. performed tr-ARPES experiments and data analysis, V. G., G. S. and E. P.  did the Bethe Salpeter calculations, L.P. conceived the experiment, wrote the article and assured funding, K.Z. and T.P. provided the sample. E. D., S. M., A. T. and M. M. contributed to the funding. All co-authors contributed to the scientific discussions and redaction of the manuscript.

\section*{Competing interests}

The authors declare no competing interests.

\bibliographystyle{naturemag}
\bibliography{Exceptional_B}


\newpage

\section*{Figures}

\begin{figure}[htp]
\includegraphics[width=\columnwidth]{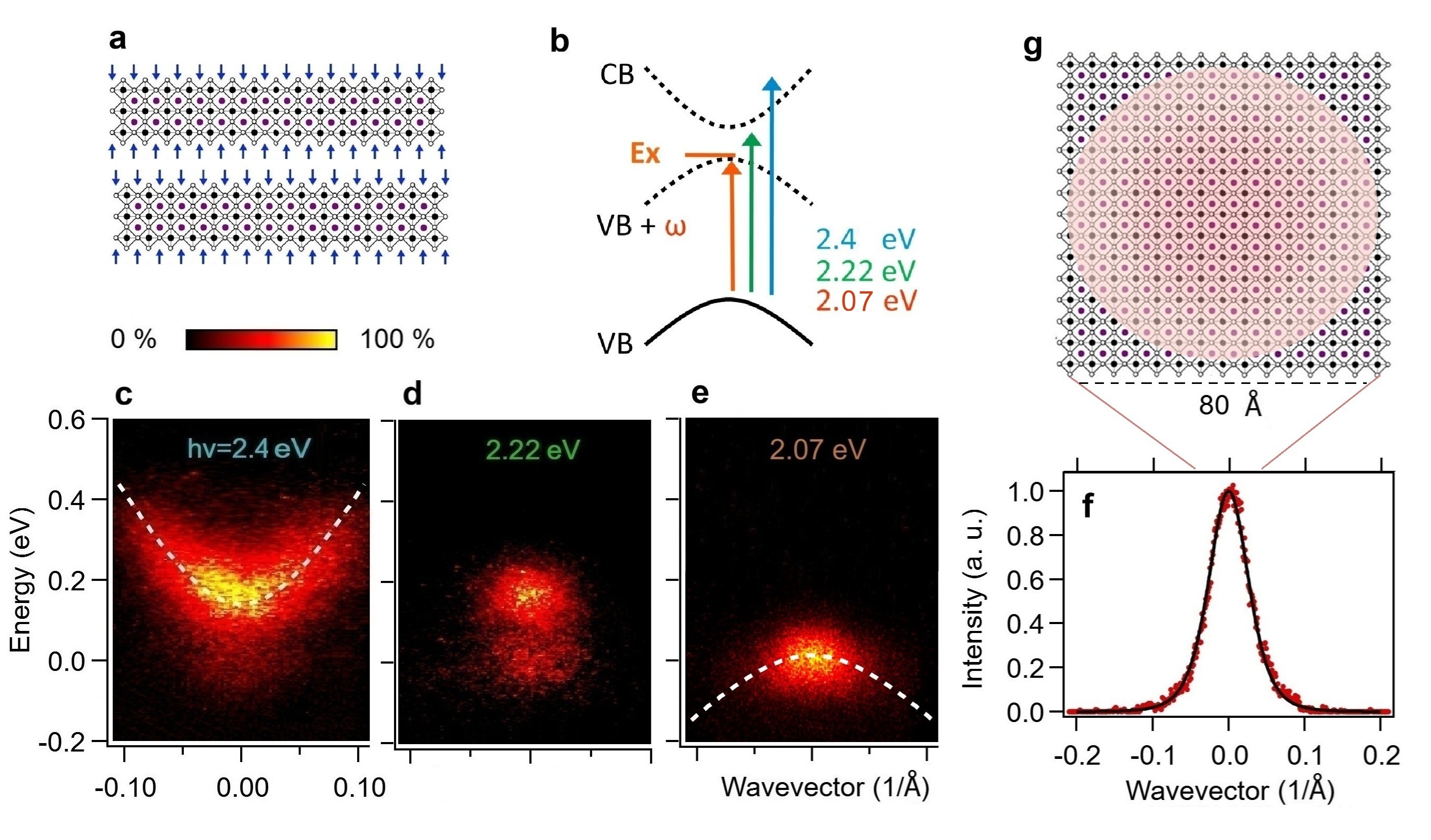}
\caption{(a) Side view of the inorganic layers and organic spacers (blue arrows) in the 2D halide perovskite (BA)$_2$(MA)$_2$Pb$_3$I$_{10}$. (b) Schematic illustration showing the optical transitions, the conduction band (CB), the valence band (VB), and its Floquet replica (VB+$\omega$). (c-e) tr-ARPES intensity maps acquired using weak pump pulses ($\sim 3~\mu$J/cm$^2$) with photon energies above the conduction-band minimum (2.4 eV, c), close to the conduction-band minimum (2.22 eV, d), and resonant with the exciton transition (2.07 eV, e). The photoelectron energy is referenced to the exciton resonance, with the conduction-band minimum located 0.18 eV above it. The dashed lines in (c) and (e) indicate the \emph{ab initio} conduction band and valence-band replica, respectively. (f) Momentum distribution curve obtained by integrating the tr-ARPES map in (e) over a 0.1 eV energy window centered on the exciton peak. The black curve shows $|\psi(k)|^2$ calculated from the Bethe–Salpeter equation in reciprocal space. (g) Real-space extent of the Wannier exciton (shaded red region), superimposed on a top view of the inorganic layer.}
\label{Exciton}
\end{figure}

\begin{figure}[ht]\
\includegraphics[width=0.8\columnwidth]{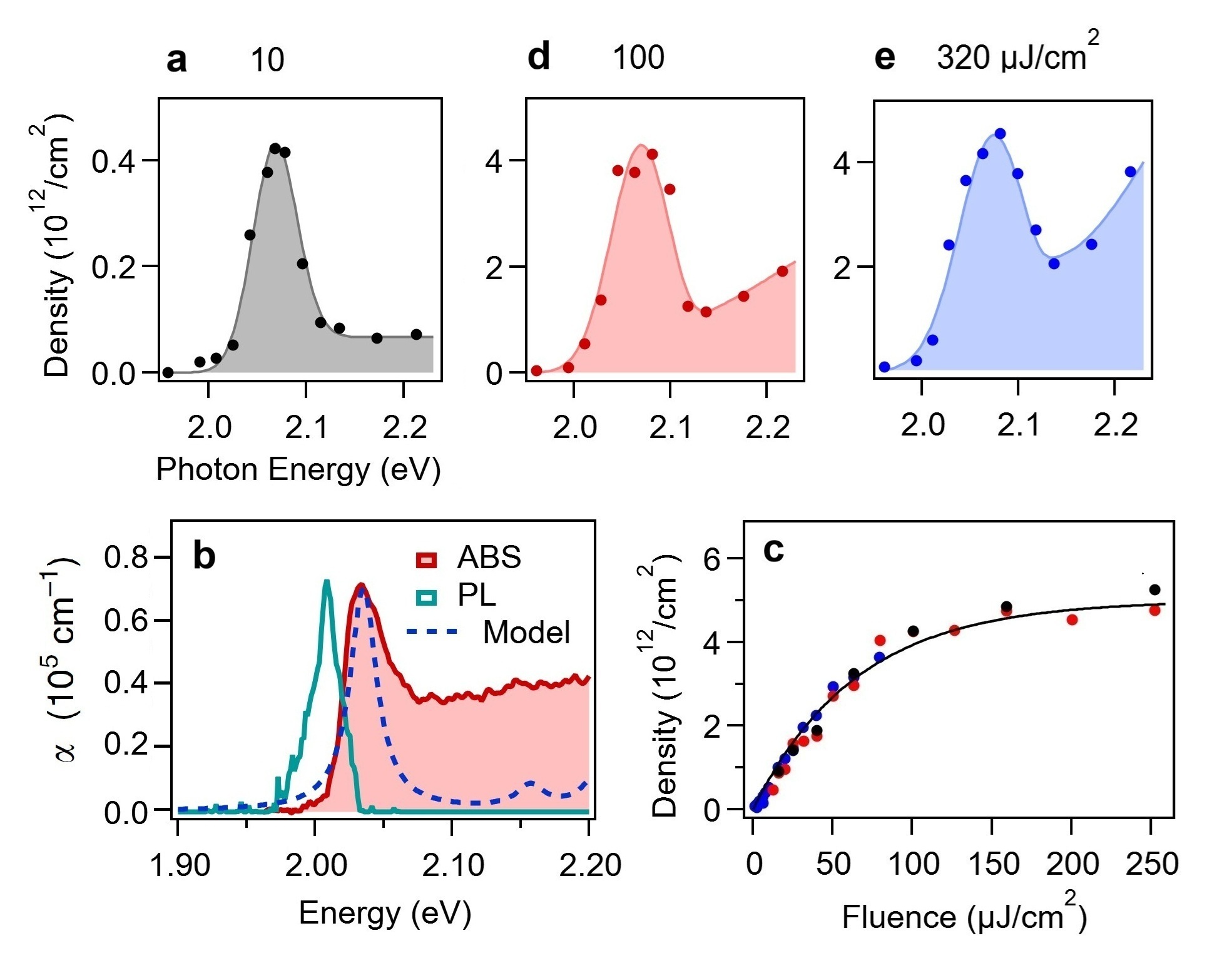}\
\caption{(a) Integrated tr-ARPES intensity at a delay time of 100 fs, measured as a function of pump photon energy at a pump fluence of 10 $\mu$J/cm$^2$. (b) Optical absorption and photoluminescence spectra. The blue dashed line shows the imaginary part of the dielectric function calculated using the Bethe–Salpeter equation and arbitrarily rescaled to match the optical absorption. (c) Integrated tr-ARPES intensity measured at a delay time of 100 fs under resonant excitation (pump photon energy of 2.07 eV) as a function of pump fluence. (d,e) Integrated tr-ARPES intensity at a delay time of 100 fs, measured as a function of pump photon energy at pump fluences of 100 $\mu$J/cm$^2$ (d) and 320 $\mu$J/cm$^2$ (e).}
\label{Resonance}
\end{figure}\

\begin{figure}[ht]\
\includegraphics[width=\columnwidth]{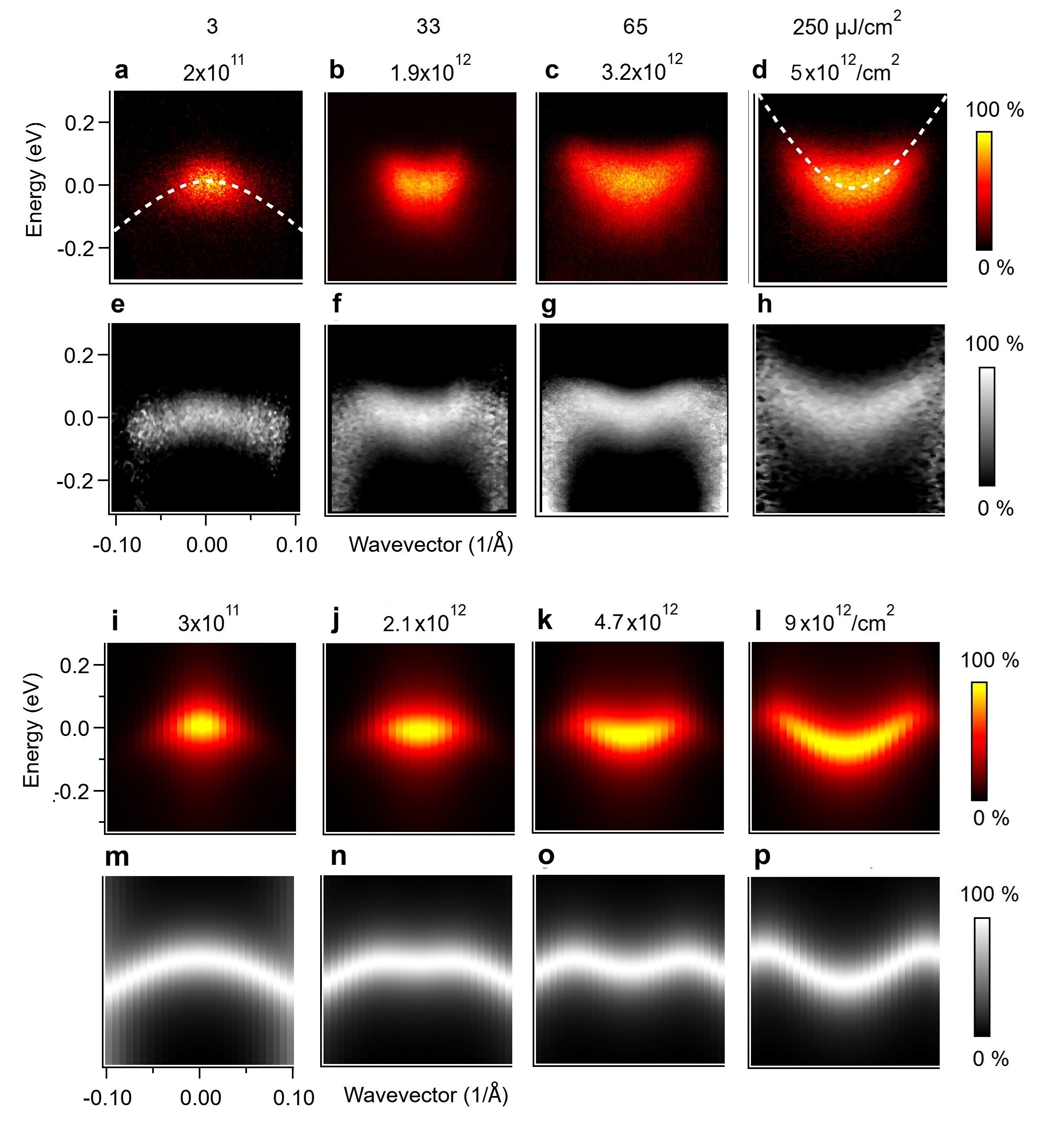}\
\caption{(a–d) tr-ARPES intensity maps acquired at delay time of 100 fs, under resonant excitation (pump photon energy of 2.07 eV), with the pump fluence increasing from 3 $\mu$J/cm$^2$ (a) to 250 $\mu$J/cm$^2$ (d). The dashed lines in (a) and (d) indicate the valence-band replica and conduction band, respectively. (e–h) Photoelectron intensity maps in which the energy distribution curve (EDC) at each wavevector is normalized to its maximum intensity. This data processing enables a clearer visualization of the spectral-weight dispersion across the entire wavevector range. (i–l) Photoelectron intensity maps modeled within the framework of the pseudo-Hermitian Bethe–Salpeter Hamiltonian (PH-BSH) under resonant pumping conditions, for photoexcitation density from $3\times 10^{11}$ (i) to $9\times 10^{12}$/cm$^2$ (l). (m–p) Maps in which the energy distribution curve (EDC) at each wavevector is normalized to its maximum intensity.
}
\label{Fluence}
\end{figure}\

\begin{figure}[ht]\
\includegraphics[width=\columnwidth]{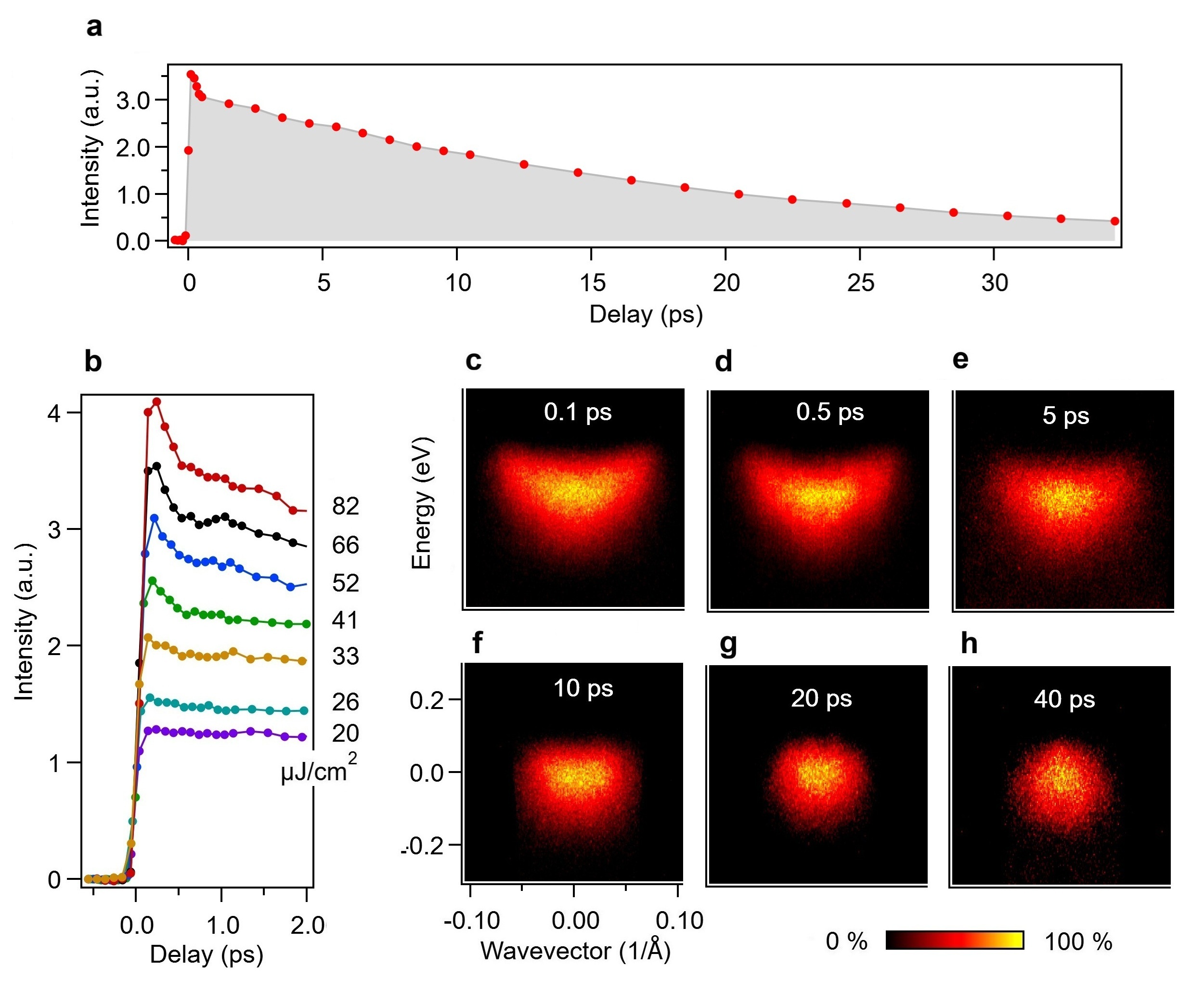}\
\caption{(a) Integrated tr-ARPES intensity acquired at a pump photon energy of 2.07 eV and a fluence of 63 $\mu$J/cm$^2$ as a function of pump–probe delay. (b) Short-time dynamics of the integrated tr-ARPES intensity for different pump fluences. The ultrafast intensity drop becomes discernible above a pump fluence of approximately 30 $\mu$J/cm$^2$. (c–h) tr-ARPES intensity maps acquired at a pump photon energy of 2.07 eV and a fluence of 65 $\mu$J/cm$^2$ at selected pump–probe delays.}
\label{Dynamics}
\end{figure}\

\end{document}